\documentclass{ws-procs975x65}

\begin{document}

\title{Killing Tensors and Conserved Quantities of a Relativistic Particle \\in External Fields}
  
\author{$^1$Takahisa Igata$^*$, $^2$Tatsuhiko Koike$^\dagger$ and $^1$Hideki Ishihara$^\ddagger$}
\address{$^1$Department of Mathematics and Physics,\\
 Graduate School of Science, Osaka City University,
 Osaka 558-8585, Japan\\
$^*$E-mail: igata@sci.osaka-cu.ac.jp\\
$^\ddagger$E-mail: ishihara@sci.osaka-cu.ac.jp\\
}

\author{}
\address{$^2$Department of Physics, Keio University, Yokohama 223-8522, Japan\\
$^\dagger $E-mail: koike@phys.keio.ac.jp}

\renewcommand{\include}[1]{}
\renewcommand\documentclass[2][]{}

\newcommand{\stp}[2]{#1 \  \stackrel{{\rm S}}\otimes \  #2}
\newcommand{\K}[1]{\stackrel{(#1)}K}
\newcommand{\T}[2]{\stackrel{(#2)}#1}
\newcommand{\pbr}[2]{\bigl\{ \, #1 \ ,\ #2 \, \bigr\} _{{\rm P}}}
\newcommand{\sbr}[2]{\bigl[ \, #1 \ ,\ #2 \, \bigr] _{{\rm S}}}

\begin{abstract}
We generalize Killing equations to a test particle system which is subjected to external force. 
We relax the conservation condition by virtue of reparametrization invariance of a particle orbit. 
As a result, we obtain generalized Killing equations which have hierarchical structure on the top of which a conformal Killing equation exists. 
\end{abstract}


\bodymatter

\section{INTRODUCTION}
Analysis of black hole physics requires a knowledge of orbits of a test particle because it gives us information of fields around a black hole. 
When we study a particle motion in a spacetime, constants of motion, conserved dynamical quantities along a particle orbit, are useful to understand the physical properties of the particle system and to integrate the equation of motion.

Here we concentrate on relationship between constants of motion of a particle and geometrical quantities such as Killing fields on a spacetime. 
We discuss generalization of Killing equation to a particle system which is subjected to external force in the Hamiltonian formalism. 
Our basic idea is a relaxation of the conservation condition for constants of motion by using reparametrization invariance of a particle orbit.

\section{FORMALISM}
Let us formulate generalized Killing equation for a general particle system which is subjected to external force in a spacetime 
$({\cal M}, g_{\mu \nu})$. 
Suppose that the general form of Hamiltonian of the system is given by 
\begin{eqnarray}
	H=\frac{N}{2m}{\cal H}
	=\frac{N}{2m} \left( g^ {\mu \nu}p_{\mu} p_{\nu} 
		+ B^{\rho}p_{\rho}+V \right),
\label{General_Hamiltonian}
\end{eqnarray}
where $m$ is mass of a particle, and $g_{\mu \nu}, B^{\rho}$ and $V$ are metric, vector field, and scalar field on ${\cal M}$, respectively. An arbitrary function $N$ of $\tau$ which parametrizes the particle orbit is a Lagrange multiplier which is related to the reparametrization invariance of the orbit.
By setting differentiation of the action $S$ by $N$ equal to zero, 
we obtain a constraint equation 
$ {\cal H}= g^ {\mu \nu}p_{\mu} p_{\nu} + B^{\rho}p_{\rho}+V  \approx 0$, 
 which defines a constraint surface. 
The weak equality $\approx$ means that the real particle motion is confined on the surface in the phase space.

If $F$, a dynamical quantity of the particle system, is a constant of motion, then $F$ and $H$ should be commutable with the Poisson bracket. 
It is adequate to satisfy the commutation relation only on the constraint surface because the particle motion is confined on the surface. 
Therefore the conservation condition becomes 
$\pbr{F}{{\cal H}}\approx0$ or that is rewritten as
\begin{eqnarray}
\pbr{F}{{\cal H}} + \phi {\cal H} = 0, 
\label{eq:condition}
\end{eqnarray}
where $\phi$ is an arbitrary function on the phase space. 
Now we assume that $F$ and $\phi$ are given as polynomial forms in canonical momenta as 
$ F  = \sum_{k=0}^{N} \stackrel{(k)}K{}^{\mu _{1} \cdots \mu_{k}}
	p_{\mu_1}\cdots p_{\mu_{k}}$
, and 
$ \phi
 = \sum_{l=0}^{M} \stackrel{(l)}\lambda{}^{\mu _{1} \cdots \mu_{l}}
	p_{\mu_1}\cdots p_{\mu_{l}}$
, where 
$\stackrel{(k)}K{}^{\mu _{1} \cdots \mu_{k}}$ 
and 
$\stackrel{(l)}\lambda{}^{\mu _{1} \cdots \mu_{l}}$ 
are symmetric tensor fields of rank $k$ and $l$ on the spacetime. 
Substituting the expressions into the conservation condition \eqref{eq:condition}, then we obtain the relation for the coefficients of canonical momenta in the form
\begin{equation}
\sbr{\K{k-1}}{\T{g}{2}}
	+
\sbr{\K{k}}{\T{B}{1}}
	+
\sbr{\K{k+1}}{\T{V}{0}}
	-
\stp{\T{\lambda}{k-2}}{\T{g}{2}}
	-
\stp{\T{\lambda}{k-1}}{\T{B}{1}}
	-
\T{\lambda}{k} \T{V}{0} =0, 
\end{equation}
where the bracket with the subscript ${\rm S}$ is Schouten bracket, and $\stackrel{{\rm S}}{\otimes}$ denotes symmetric tensor product.
We note that the highest-order equation is a conformal Killing equation of rank-$N$. 
Since the equations have a hierarchical structure, we call them {\it hierarchical equations}. 
The hierarchical equations are understand as a generalization of Killing equation.

\section{APPLICATION}
Now we apply the formalism of the hierarchical equations to a dynamical system of a charged particle. 
The Hamiltonian of the system is in the case of $B^{\mu} = - 2 q A^{\mu}$ and $V = q^2 A^{\mu}A_{\mu} + m^2$, 
where $q$ is electric charge of a test particle, and $A_{\mu}$ is a gauge potential on a spacetime. 
We pay attention to the case of $N=2$ because we are interested in quadratic constants of motion in almost case when we solve equations of motion.
The corresponding hierarchical equations are written by 
\begin{eqnarray}
&&
- \sbr{\K{2}}{g} 
 + 
\stp{\T{\lambda}{1}}{g} 
 =0,
\nonumber\\
&&
- \sbr{\K{1}}{g} 
 + 
2q \sbr{\K{2}}{A} 
 +
\T{\lambda}{0} g 
 -
2  q  \stp{\T{\lambda}{1}}{A} 
 =0,
\nonumber\\
&&
- \sbr{\K{0}}{g} 
 + 
2 q \sbr{\K{1}}{A} 
 -
q^2 \sbr{\K{2}}{A^2}
 -
2 q \T{\lambda}{0} A 
 +
\T{\lambda}{1}(q^2 A^2 + m^2) =0.
\end{eqnarray}
The consistency of the hierarchical equations requires the relation 
\begin{eqnarray}
m^{2} (  \T{\lambda}{0} + q  \T{\lambda}{1}{}_{\mu}  \T{A}{1}{}^{\mu}  )
 =0. 
\end{eqnarray}
It is worth noting that the highest-rank equation remains as a conformal Killing equation. 
There exist a constant of motion of a charged particle if the hierarchical equation associated with a conformal Killing tensor field admits a nontrivial solution. 
The resultant quantity would be a constant of motion conserved only on the constraint surface in a phase space.

Then, we apply the hierarchical equations to a charged particle system in explicit backgrounds. 
First example is a constant of motion of a charged particle in a test electromagnetic field which is a solution of vacuum Maxwell's equations on a Kerr background. 
As shown by Wald \cite{Wald:1974np}, 
the gauge potential given by 
$
A^{\mu} = c_1  \xi^{\mu}  + c_2  \psi^{\mu}
\label{Wald's_sol}
$
 solves vacuum Maxwell's equation, where $\xi$ and $\psi$ are the stationary and the axial Killing vector fields, and $c_1$ and $c_2$ are constants. 
Since the metric admits a Killing tensor field, which solves the top of the hierarchical equation, we consider the case of $N=2$. 
By inspecting the integrability condition of the hierarchical equations, 
we find that this equation is integrable only when $c_2=0$. 
In the integrable case, $A^{\mu}=\xi^{\mu}$, the set 
\begin{eqnarray}
	\K{2}{}_{\mu \nu}=\gamma K_{\mu \nu }, \quad
	\K{1}{}^{\mu}=\alpha \xi^{\mu}  + \beta \psi^{\mu},  \quad
	\K{0} = \gamma K_{\mu \nu}\xi^{\mu} \xi ^{\nu}
\end{eqnarray}
is a solution, where $\alpha$, $\beta$, and $\gamma$ are arbitrary constants. 
Finally, a constant of motion of a charged particle associated with rank 2 tensor is given by
$
F= \gamma ( K^{\mu \nu}p_{\mu} p_{\nu} + K_{\mu \nu}\xi^{\mu} \xi ^{\nu})
+( \alpha \xi ^{\mu} + \beta \psi^{\mu}) p_{\mu}.
$
Therefore we obtain the constants of motion in canonical momenta 
$\xi ^{\mu} p_{\mu}$, $\psi^{\mu} p_{\mu}$, and $K^{\mu \nu }u_{\mu}u_{\nu}$.

Second example is the system in a five-dimensional charged rotating black hole which is a typical solution in the five-dimensional Einstein-Maxwell-Chern-Simons theory characterized by mass parameter $m$, charge parameter $q$ and two spin parameters $a$ and $b$ \cite{Chong:2005hr}. 
By solving the hierarchical equations, we obtain the quadratic constant of motion of a charged particle moving in the five-dimensional charged Kerr black hole
 as follows 
\begin{eqnarray}
F =  K^{\mu \nu} p_{\mu} p_{\nu} + 2 S A^{\mu} p_{\mu} -  S A^{\mu}A_{\nu}. 
\end{eqnarray}
where $S = a^2 \cos ^2 \theta + b^2 \sin ^2 \theta$ is a metric function. 

These examples are in the case of $\T{\lambda}{1}=0$. 
That is, constants of motion are associated with Killing tensor fields. 
We note that we can show examples for $\T{\lambda}{1}\neq0$, artificially, {\it i.e.}, a constant of motion associated with conformal Killing tensor field. 
It will be reported in a forth coming paper.

\bibliographystyle{ws-procs975x65}

\end{document}